\documentclass[10pt,a4paper,twocolumn]{article}
\usepackage{paper_style}
\usepackage[activate={true,nocompatibility},final,tracking=true,kerning=true,spacing=true,factor=1100,stretch=10,shrink=10]{microtype}
\usepackage{amsthm}
\usepackage{capt-of}
\usepackage{makecell}
\usepackage{tikz}
\usepackage{esvect}

\usepackage{subcaption}
\usepackage{graphicx}
\usepackage{adjustbox}
\usepackage{dblfloatfix}

\makeatletter
\newcommand{\thickhline}{%
    \noalign {\ifnum 0=`}\fi \hrule height 1pt
    \futurelet \reserved@a \@xhline
}
\newcolumntype{"}{@{\hskip\tabcolsep\vrule width 1pt\hskip\tabcolsep}}
\makeatother
\title{\vspace{-0.8in}{\bfseries 
    \Large{Synthetic vascular structure generation for unsupervised pre-training in CTA segmentation tasks}}\\
}
\author{
  \textbf{Nil Stolt Ans\'o}\\
  University of Amsterdam\\
  \texttt{nilstoltanso@gmail.com}
}

\date{\vspace{-5ex}}

\begin{document}

\twocolumn[
  \maketitle
  \begin{@twocolumnfalse}
    \begin{abstract}
Large enough computed tomography (CT) data sets to train supervised deep models are often hard to come by. One contributing issue is the amount of manual labor that goes into creating ground truth labels, specially for volumetric data. In this research, we train a U-net architecture at a vessel segmentation task that can be used to provide insights when treating stroke patients. We create a computational model that generates synthetic vascular structures which can be blended into unlabeled CT scans of the head. This unsupervised approached to labelling is used to pre-train deep segmentation models, which are later fine-tuned on real examples to achieve an increase in accuracy compared to models trained exclusively on a hand-labeled data set.
{ \vspace{6ex}}

\end{abstract}
  \end{@twocolumnfalse}
]

\thispagestyle{firststyle}
\setcitestyle{square}
\bibliographystyle{unsrt}

\section{Introduction}
Angiography is a visualization technique whereby injecting a dye into the circulatory system an image of the inside of blood vessels can be obtained. A common method through which to obtain angiographic data is computed tomography (CT). A variety of fields have benefited from recent advances in CT angiography (CTA), one of them being the treatment of acute ischemic stroke. 

Given that for stroke patients the elapsed time since the stroke onset is critical to recovery, fast and precise imaging of affected brain regions is crucial. A common first step in the diagnosis of affected tissue is vessel segmentation. A bottleneck to this process is in the localization and evaluation of regions of interest, which often relies on an emergency radiologist having to manually go over the volumetric data. It comes to no surprise that automation techniques dedicated to speeding up this process have been seeing significant development in recent years.

\subsection*{Imaging Techniques}
A variety of methods exist to capture volumetric angiographic information. X-ray based methods like computed tomography angiograhy (CTA) tend to be more commonplace over magnetic resonance angiography (MRA). Technical reasons for this include higher spatial resolution, although there are also practical reasons such as not having to check if the patient is susceptible to MR.

Image processing techniques have been developed to segment target regions in volumetric data. However, the design of robust techniques is non-trivial. There are structures in the cranial cavity with equal intensity values as the contrast material which often result in false positives under naive segmentation approaches. Furthermore, determining the exact boundaries of vessels that are located close to bone structure is a challenge of its own. Methods in clinical use tend to revolve around the use of an atlas, which base decisions given density models created from manually annotated examples~\cite{atlas1, atlas2, atlas3}. However, atlas based methods are limited in their ability to generalize to variability in individual patient anatomy.

\subsection*{Deep Learning}
With the increase of annotated data, supervised learning techniques have in recent years become prevalent in medical imaging fields. Convolutional neural networks (CNNs) have shown great promise in diagnosis from visual information. In recent studies, these approaches have began to surpass domain experts in terms of early detection and overall accuracy~\cite{breast_cancer}. With enough data, these networks have in some cases even managed to discover underlying correlations previously unknown to experts \cite{iris}. It thus comes with no surprise that there is a growing incentive to make use of the apparent ability of these models to discover and utilize features in the data that are unperceivable to humans.

One obvious downside to deep models is the amount of data required to tune their parameters. Commonly used CNNs tend to have numbers of parameters in the order of hundreds of thousands, if not millions. However, access to large amounts of data in many medical imaging fields is often impossible. One obvious reason to this scarcity is rarity of certain conditions. However, other roadblocks are present. Patient anonymity concerns make it impossible for data sets to be compiled and distributed. Furthermore, data labelling entails lots of tedious manual labour. This is worsened by the fact that domain experts are commonly the only ones qualified to perform such tasks.

\subsection*{Contributions}
This paper aims to explore data efficiency techniques that can prove useful for training a deep model for brain vessel segmentation tasks in CTA scans.

First, we propose a synthetic vessel generation algorithm than can be used to pretrain a CNN model. The generated vascular structure aims to mimic the appearance of the arterial system observed in real CTA scans. This allows the use of unlabeled CTA scans by blending generated synthetic vascular structures into them.

Second, we explore whether the recreation of the underlying noise in the scans can serve as a form of augmentation. The process consists on denoising a scan and applying noise similar in nature to CT noise.

\section{Previous research}

\subsection*{Synthetic Vessel Generation}

The arterial system of the human body is a complex branching structure used to deliver blood from the heart to different parts of the body. The physiology of arterial trees has evolved, over the course of millions of years, to efficiently supply tissue with oxygen and nutrients, as well as removing metabolic end products~\cite{karch2000staged}.

First attempts to incorporate these principles into discrete computational models include~\cite{2d_vessels_1991, vascular_trees, 2d_vessels_1996}. The produced 2-D vascular structures by their early angiogenesis models, although milestones at the time, were artificial and simplistic in appearance.

With an increase in biological data, newer computational models of arterial trees have been developed in a range of research fields, each accounting for influencing variables at different size scales. Approaches such as the one taken by~\cite{karch2000staged} and~\cite{schneider2012tissue} aim to account for macroscopic influencers such as oxygen concentration in surrounding tissue and fluid flow resistance within the generated vessel structure. Other approaches such as~\cite{milde2008hybrid} attempt to simulate microscopic influencing variables such as the formation of endothelial tip cells and the influence that proteins and hormone densities in the extracellular matrix have on tip cell trajectories.

A radically different approach involves generating vascular structures using a learned distribution from a data set. A common choice for such tasks are Generative Adversarial Networks (GAN)~\cite{GAN}. Although these methods are only capable of creating an implicit model of the data they are presented, they have been shown to produce anatomically convincing vascular structures~\cite{retina_GAN, retina_GAN_segmentation}.

\subsection*{Vessel segmentation}
Early approaches to vessel segmentation tended to revolve around the application of unsupervised computer vision techniques. A common approach to help highlight vascular structures involves multi-scale analysis, whereby local second-order information (Hessian) is examined~\cite{hessian}. A similar approach is that of matched filtering~\cite{staal2004ridge, retina_vesselseg}, where a variety of Gaussian based filters (such as Gabor filters) are convolved with the image to enhance features. Another example of a used approached is that of vessel tracking. This approach involves tracing a vessel based on local structure~\cite{region_growing}. Vessel tracking however requires root positions to be predefined.

In recent years, supervised approaches have shown dominance over the aforementioned methods thanks to their ability to make use of ground truth data. Convolutional neural networks (CNN) have shown great promise due to their hierarchies of features learnt directly from the data. Approaches such as~\cite{retina_CNN, retina_CNN2} were among the first to show the potential of CNNs to vessel segmentation tasks on retina images. Following advances in 3D CNNs,~\cite{tetteh2018deepvesselnet} proposed approaches to efficiently treat large CT and MR scan volumes for vessel segmentation of the head. These included replacing 3D kernels by 1D cross-hair filters and pretraining a model on synthetic vessel data. More recent state-of-the-art research on volumetric data has made use of U-Net architectures~\cite{U-net}. Skip-connections in the architecture allow for local information to be preserved across feature extraction layers. This makes the U-nets ideal for localization of small anatomical structures~\cite{DBLP:journals/corr/abs-1809-04430, roth2018application}.

\section{Methodology}
\subsection*{Synthetic Vessel Generation}
The synthetic vessel generation approach employed in this research was in many ways similar to the rooted tree branching structure found in~\cite{schneider2012tissue}. A 3D tree structure grows from a root node, where new inter nodes can grow from current leaf nodes, and inter nodes can spawn new leaf nodes to form bifurcations. The core idea behind our approach is to generate two trees such that each root node is approximately located at the coordinates of different internal carotid artery so as simulate a synthetic arterial system branching into the brain.

In a given tree, each node has a three-dimensional position vector $\vec{p}$ and a direction vector $\vec{d}$. For all nodes except for the root node, the starting position vector $\vec{p}$ is the tip of its parent's direction vector ($\vec{p}_p + \vec{d}_p$). The root node's position vector $\vec{p}$ is a predefined coordinate given by a internal carotid's average entry point to the brain given by the subjects in our data set. The direction vector of the root node was set to point straight up so as to have the vessel structure grow upwards into the brain (since the root's position is in the base of the brain). The direction vector of a child node $\vec{d}_c$ depended on whether the node's parent was a leaf-node or if the new node was a bifurcation. 

In the simple case that a child node was spawned from a leaf node, the child's direction vector is initialized to be equal to it's parent ($\vec{d}_c = \vec{d}_p$) and subsequently perturbed in two ways. The first perturbation is performed by adding Gaussian noise $N(0, \sigma)$ to the angle component of the polar coordinates, where $\sigma$ was empirically set to $\pi/32$ radians. The second perturbation is in the direction of a sampled target point in order to guide the leaf node towards points of interest (more on this later). The post-perturbation vector $\vec{d_c^*}$ is always adjusted in magnitude so as to satisfy $\|\vec{d}_c\| = \|\vec{d}_p\|$. This will be ensured by modifying $\vec{d_c^*}$ in the following manner: 
\begin{equation}
    \vec{d}_c = \|\vec{d_p}\| \frac{\vec{d_c^*}}{\|\vec{d_c^*}\|}
    \label{inter_node_radius}
\end{equation}

In the case that the child node was a product of a bifurcation, the direction vector was determined similar to~\cite{schneider2012tissue}, where the bifurcation angle is calculated in accordance to fluid dynamics described in~\cite{Fung_Biomechanics} and the minimum work and energy dissipation proposed in~\cite{Murray299} and~\cite{Optimality_principles}:
\begin{equation}
    \cos \left(\phi_{c}\right)=\frac{r_{p}^{4}+r_{c}^{4}-r_{e}^{4}}{2 r_{p}^{2} r_{c}^{2}}
    \label{child_angle}
\end{equation}
where $\phi_{c}$ is the bifurcation angle of the new child, and $r_{p}$, $r_{c}$, $r_{e}$ are the radius of the parent, new child, and existing child nodes respectively. 

\begin{figure}[b!]
    \centering
    \includegraphics[width=\columnwidth]{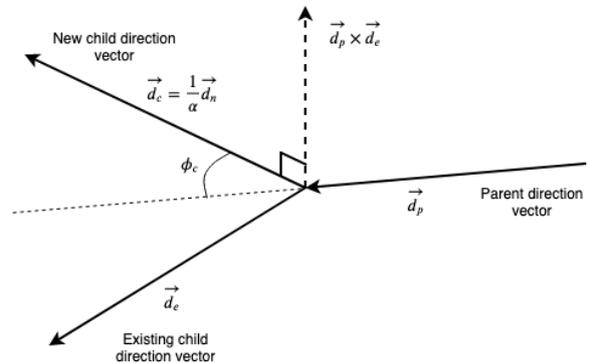}
    \caption{Representation of direction vectors when a new child is created in a bifurcation node.}
    \label{fig:child_rot}
\end{figure}

The new child's direction vector $\vec{d}_c$ is coplanar to $\vec{d}_p$ and $\vec{d}_e$. Once the angle is obtained using eq.~\ref{child_angle}, a new direction vector $\vec{d}_n$ is calculated using Rodrigues' rotation formula, where a copy of $\vec{d}_e$ is rotated by ($\phi_{c} + \phi_{e}$) around the pivot vector $\vec{d}_p\times\vec{d}_e$ that is normal to the plane containing $\vec{d}_p$ and $\vec{d}_e$ (Figure~\ref{fig:child_rot}). The resulting $\vec{d}_n$ has magnitude $\|\vec{d}_n\| = \|\vec{d}_e\| = \alpha\|\vec{d}_c\|$, where $\alpha$ is the ratio between magnitudes of $r_c$ and $r_e$. The vector $\vec{d}_n$ has to be subsequently scaled in order to obtain $\vec{d}_c$ (more on this below).

\begin{figure}[h]
  \begin{subfigure}{0.49\columnwidth}
    \includegraphics[width=\linewidth]{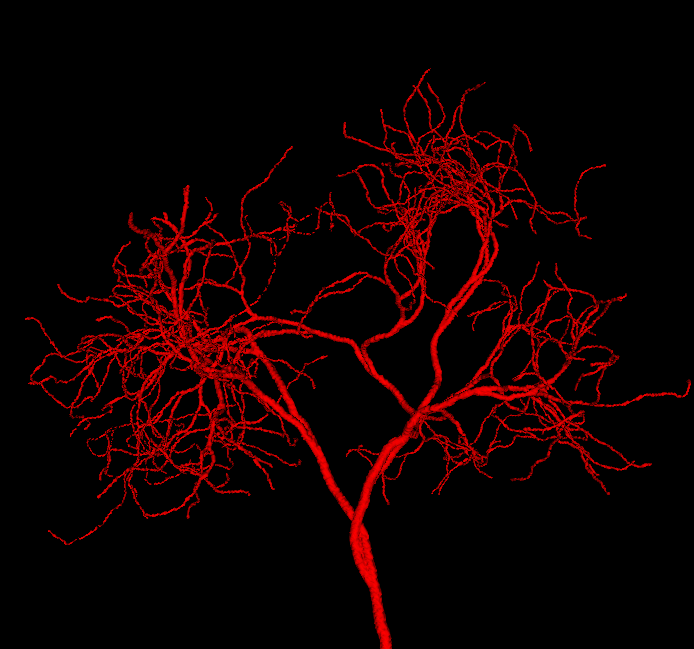}
  \end{subfigure}
  \hfill %%
  \begin{subfigure}{0.49\columnwidth}
    \includegraphics[width=\linewidth]{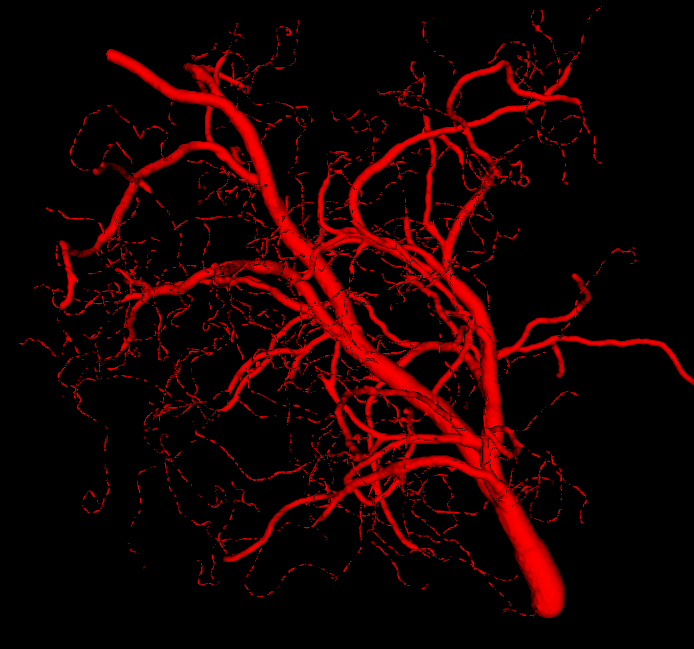}
  \end{subfigure}
  \caption{Generated vessel structures with no spatial target growth incentive. The right figure was grown in a small spatial volume relative to the root width, while the left was grown unconstrained by volume.}
  \label{fig:trees}
\end{figure}

The radius of the vessel is tied to the direction vector of it's nodes. Each node is represented as a sphere with center at a given node's position vector $\vec{p}$ and with radius proportional to the direction vector magnitude $r = \beta\|\vec{d}\|$. With enough overlap between spheres ($\beta > 1.0$), a tube-like structure is approximated. The binary vessel structure can be obtained in 3-dimensional pixel space through rasterization of the tree nodes into voxel space. The value of $\beta$ not need be much larger than $1.0$ for the rendered vessels to appear realistic, as the resolution of a CTA scan is always too low compared to the size scale of the imperfections arising from this approximation. We empirically set a node's starting $\beta$ value to 3.0, as even in branches that had their radii reduced due to bifurcations taking place in earlier nodes (while having parent-child distances preserved), no visible width disturbances could be observed (see Figure~\ref{fig:trees}).

Given that we determine a node's radius to be $r = \beta\|\vec{d}\|$, if we know the radius of a bifurcation's new child $r_c$, the direction magnitude $\|\vec{d}_c\|$ can be obtained. The vector $\vec{d}_c$ is the calculated by scaling the normalized $\vec{d}_n$ vector by $\|\vec{d}_c\|$: 

\begin{equation}
    \vec{d}_c = \frac{\vec{d}_n}{\|\vec{d}_n\|} \frac{r_c}{\beta}
    \label{bifurcation_direction}
\end{equation}

The radius of a new child node also depends on whether the parent is a leaf node becoming an inter node or an inter node becoming a bifurcation. If the parent node is a leaf node, the new child will be assigned the same $\beta$ as the parent so as to preserve the same radius along the vessel. However, in the case of a bifurcation node, the children will have radii conforming Murray’s law~\cite{Murray299}:

\begin{equation}
    r_{p}^{\gamma}=r_{l}^{\gamma}+r_{r}^{\gamma}
    \label{murray_lay}
\end{equation}
where $\gamma$ denotes the bifurcation exponent. A value of $\gamma = 3.0$ was chosen for our model, although values of $2.0 < \gamma \leq 3.0$ have been reported in the literature~\cite{schneider2012tissue, vascular_trees, canine_coronary_artery}.

To determine the radius assigned to the new child in a bifurcation, we follow the heuristic proposed in \cite{schneider2012tissue}. According to Murray's law for a fully symmetric bifurcation ($r_c = r_e$), the "expected" radius $r_\mu$ of distal segments is defined as $r_c = 2^{-1/\gamma}r_p$. The new child radius is sampled from the narrow normal distribution $\mathcal{N}\left(\mu=r_{\mu}, \sigma=r_{\mu} / 32\right)$ and clipped between $0.0<r_c<r_p$. Once $r_c$ is known, the existing child's radius is subsequently modified by rearranging equation 3.4:

\begin{equation}
    r_{e}=\left(r_{p}^{\gamma}-r_{c}^{\gamma}\right)^{1 / \gamma}
    \label{murray_lay_re}
\end{equation}

After a bifurcation, if any of the two children have a smaller radius than $r_{min}$, the child is pruned. In the case of pruning, the remaining child will then have its radius become equal as its parent $r_p$ (thus the parent becoming an inter node). If the existing child happened to be the one pruned and had children of his own, the entire child's branch is recursively removed. Similarly, if the none of the two children in the bifurcation were pruned, the existing child's branch will have its radius recursively reduced by the ratio $r_{e\_new}/r_{e\_old}$. If any of the subbranches happened to contain nodes with radius $r<r_{min}$, those subbranches are recursively removed.

\begin{figure*}[t]
  \begin{subfigure}{.66\columnwidth}
    \includegraphics[width=\linewidth]{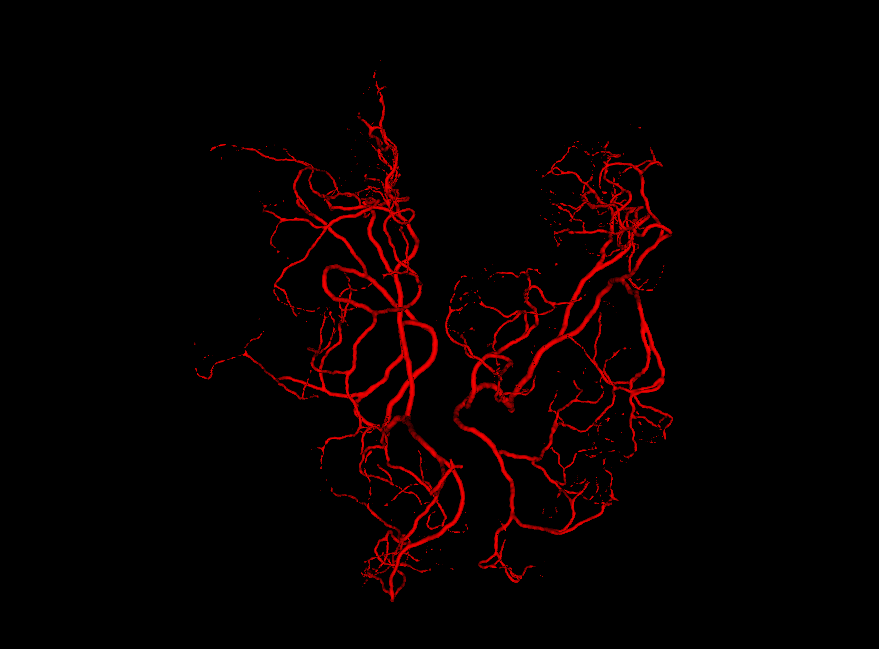}
    \caption{Axial (top-down) view.}
    \label{fig:tree_target_axial}
  \end{subfigure}
  \hfill %%
  \begin{subfigure}{0.66\columnwidth}
    \includegraphics[width=\linewidth]{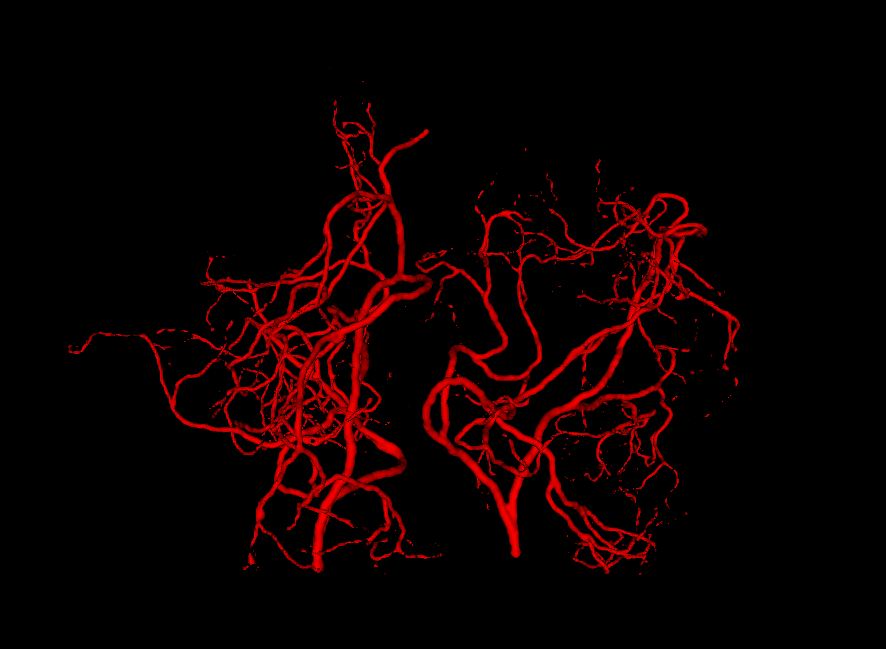}
    \caption{Coronal (back-front) view.}
    \label{fig:tree_target_coronal}
  \end{subfigure}
   \hfill %%
  \begin{subfigure}{0.66\columnwidth}
    \includegraphics[width=\linewidth]{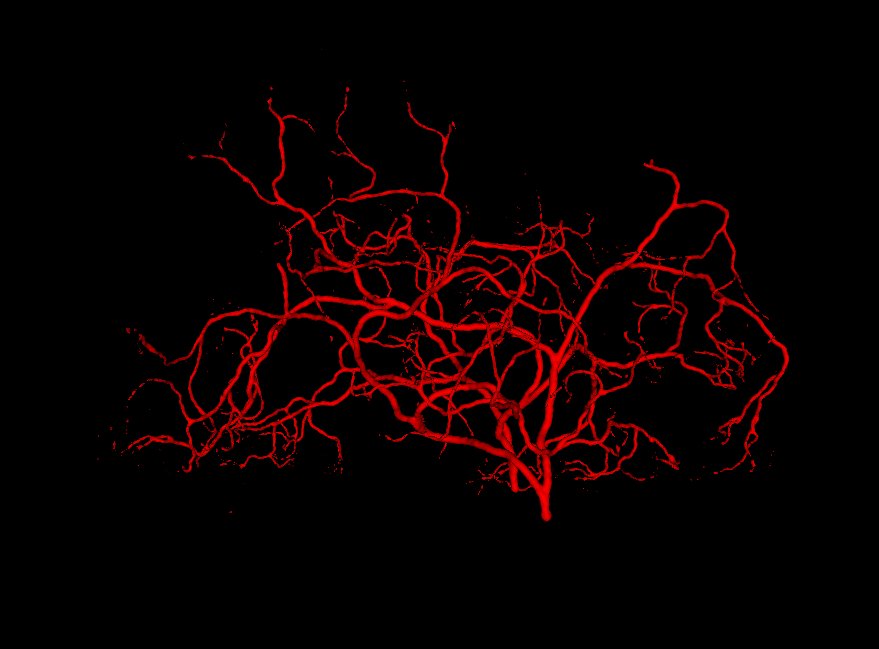}
    \caption{Sagittal (left-right) view.}
    \label{fig:tree_target_sagittal}
  \end{subfigure}
  \caption{Generated vessel structures with spatial target growth incentive. The target volume was a vessel density atlas created by averaging the ground truth of patients in our data set. }
  \label{fig:hemisphere_trees}
\end{figure*}

When growing the tree, two separate lists of nodes were kept updated. One list contained all current leaf nodes, while the other contained all inter nodes capable of undergoing bifurcation. A predefined bifurcation probability was used to determine whether the next created child node would be a product of a leaf-node becoming an inter node or an inter node becoming a bifurcation. Depending on which of the two cases was decided, a random node was sampled from the corresponding list.

The sampling was performed under a uniform distribution. Previous tests dealing with weighted sampling based on a node's region within the brain (more on this below) lead to single branches growing disproportionately while others leaf nodes never had a chance to grow. Also, certain requirements are needed to be satisfied for an inter node to be considered as capable of bifurcation. First, the radius of the node had to be at least $2r_{min}$ (otherwise one of the children was guaranteed to be pruned right away). Second, the node couldn't be a leaf node. Lastly, the node had to be no closer than a predefined number of nodes away along a vessel from an existing bifurcation.

One key point to note is that, unlike in~\cite{schneider2012tissue}, we do not back-propagate through the entire tree in order to optimize every node's angles and widths when a bifurcation takes place. Such an approach would increase the computational complexity of the algorithm significantly, and subsequently the time required to generate these structures. Our main motivation for building these vessel structure is to optimize the performance of our segmentation algorithm through pretraining, and as such, extensive improvements to realism start to bring diminishing returns. Furthermore, the resolution at which the vessels are rendered into the scans is likely to help hide minor imperfections. It is sufficient that the generated vessel structures appear to be plausible to the feature extraction layers of our convolutional network. 

An approach was also introduced to make the global vessel distribution mimic the one found in the brain. The brain arterial system is designed to supply the entire brain, yet the current structure described so far grows away from the root node and is unlikely to meaningfully fill the volume of the brain once rendered. To solve this issue, a density atlas was created out of the patients scans in our data set to assign a volume with probabilities of an artery being present. 

This volume was turned into binary values depending on whether a region had a probability density above a certain threshold. This transformation proved to be helpful to prevent all neighbouring vessels within a region to cluster into the same high-probability regions, and instead spread evenly out. When a leaf node was created (except in the bifurcation case), the surrounding region to the nodes position was sampled in the atlas. The sampled point with the highest value $\vec{s}_{max}$ was then used to create a target vector $\vec{t} = \vec{s}_{max} - \vec{p}_c$ towards which to rotate the new child's $\vec{d}_c$. After the rotation, the resulting vector is normalized to have the same length as its parent:
\begin{equation}
    \vec{d}_{c\_new} = \| \vec{d}_{p}\|\frac{\vec{d}_{c\_old} + \lambda(\vec{t} - \vec{d}_{c\_old})}{\|\vec{d}_{c\_old} + \lambda(\vec{t} - \vec{d}_{c\_old})\|}
    \label{target}
\end{equation}
where $\lambda$ determines how much to rotate $\vec{d}_{c\_old}$ towards $\vec{t}$. 

To incentivize vessels to spread out, every time a node is created the neighboring region around the node in the atlas was reduced in value. The neighboring region was in the form of a sphere around the node's position $\vec{p}_c$. The neighborhood radius was linearly proportional to the radius of the node $r_c$.

Another use of the vessel atlas is to determine if a node is out of bounds. This is done by checking if the atlas has a value of 0 at the node's position. If a leaf node is too far out of the brain, the branch is discontinued (the leaf will never be selected to spawn a child). On top of this, in order to make the overall structure more realistic, a different atlas was used for each of the two hemisphere root nodes. Such atlas would only have non-zero values for one brain hemisphere. This causes the resulting vessel structures to no longer cross over between hemispheres. A resulting example structure can be seen in Figure \ref{fig:hemisphere_trees}.

\subsection*{Data sets}
Available to us was a set of 4D CTA scans belonging to 57 stroke patients. Out of these, 24 had labeled artery ground truths, while the other 33 were unlabeled. Given that the goal of this research was to improve the segmentation on a single 3D scan, we treated each of the patient's frames in the time dimension as independent scans.

The labeled data set was one where each individual 4D scan had one single 3D ground truth. The labelling of each patient 4D scan was performed by hand by moving along the time dimension and marking the volume highlighted by the contrast material over time. The 3D time frames that made up our training, validation, and test sets were manually selected based on presence and clarity of the vessels. We explicitly aimed to use images from peak arterial, equilibrium, and peak venous phases of the 4D CTA scan. Time steps from a 4D scan's early arterial phase were manually discarded as insufficient contrast material had reached the brain and no vessels were visible. Similarly, late time steps belonging to late venous phases were discarded due to the contrast being too diffused. The final training set contained 243 3D scans originating from 20 patients. The validation set consisted of 26 3D scans originating to two different patients. Lastly, the test set consisted of 22 3D scans belonging to yet two other different patients. A given patient's set of 3D scans were purposely not split across multiple of these data sets (train, validation, and test) to prevent the model from using learned features specific to individual patient's anatomy.

For each of the selected 3D frames, a specific ground truth was created to correspond with the visible contrast in the frame. A threshold was applied to the general ground truth so as to remove segmentation of arteries not yet highlighted on a given 3D frame. Furthermore, since the human segmentation had occasional under-segmentation in some frames, the ground truth was enhanced by expanding the segmentation to neighboring pixels in the upper and lower axial slice if such pixels had intensity values within range of typical contrast values ($95 <= I < 450$).
\begin{figure*}[t]
\vspace{-1.5cm}
\centering
    \begin{adjustbox}{width=\textwidth}
    \begin{tikzpicture}{width=12cm}
    \centering
        \node[align=left] at (0,-0.2) {\includegraphics[width=2.0cm]{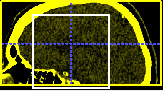}};
        \node[text width=1.85cm, align=center] at (0,0.6) {Sagittal};
        \node[align=left] at (2.2,0) {\includegraphics[width=2.0cm]{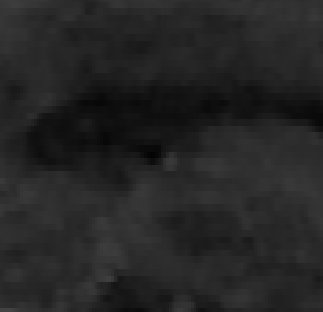}};
        \node[align=left] at (4.4,0) {\includegraphics[width=2.0cm]{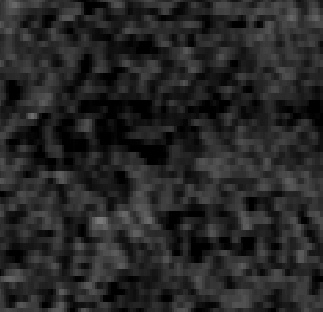}};
        \node[align=left] at (6.6,0) {\includegraphics[width=2.0cm]{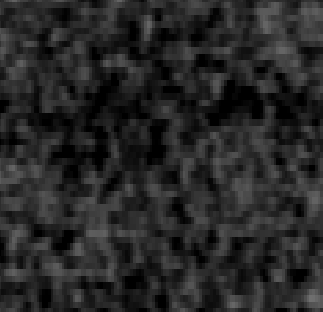}};
        
        \node[align=left] at (0,1.9) {\includegraphics[width=2.0cm]{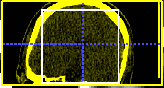}};
        \node[text width=1.85cm, align=center] at (0,2.6) {Coronal};
        \node[align=left] at (2.2,2.0) {\includegraphics[width=2.0cm]{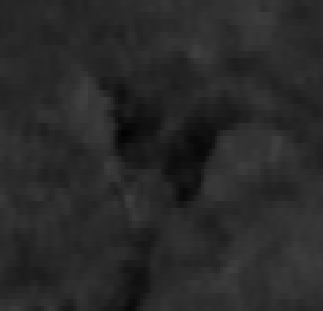}};
        \node[align=left] at (4.4,2.0) {\includegraphics[width=2.0cm]{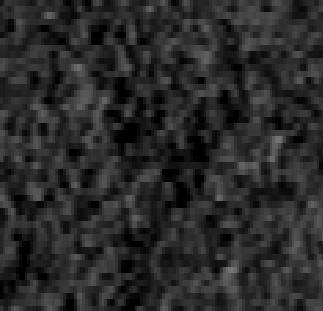}};
        \node[align=left] at (6.6,2.0) {\includegraphics[width=2.0cm]{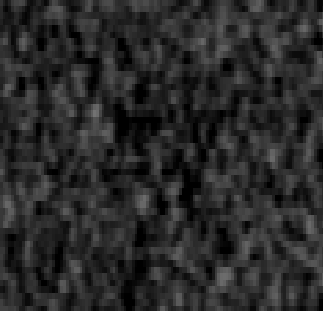}};
        
        \node[align=left] at (0,3.9) {\includegraphics[width=1.5cm]{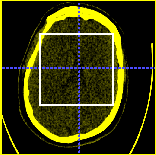}};
        \node[text width=1.85cm, align=center] at (0,4.8) {Axial};
        \node[align=left] at (2.2,4.) {\includegraphics[width=2.0cm]{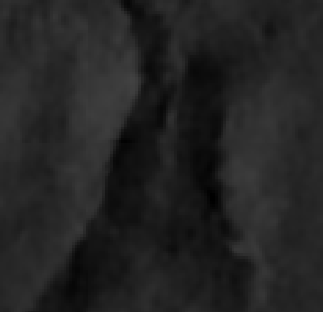}};
        \node[align=left] at (4.4,4.) {\includegraphics[width=2.0cm]{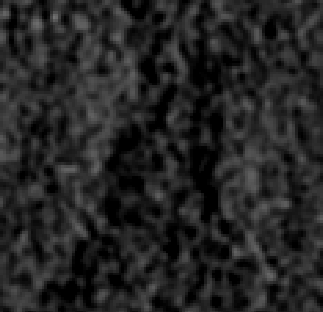}};
        \node[align=left] at (6.6,4.) {\includegraphics[width=2.0cm]{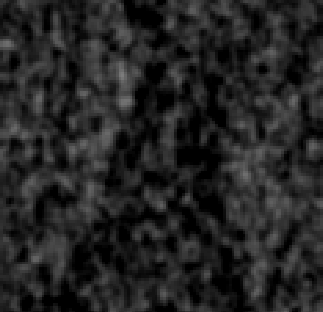}};
        
        \node[text width=2.0cm, align=center] at (2.2,5.2) {\small{Denoised}};
        \node[text width=2.0cm, align=center] at (4.4,5.5) {\small{Generated }};
        \node[text width=2.0cm, align=center] at (4.4,5.2) {\small{Perlin noise}};
        \node[text width=2.0cm, align=center] at (6.6,5.2) {\small{Original noise}};
    \end{tikzpicture}
    \end{adjustbox}
    \caption{CTA slices showing the lateral ventricles. The maps on the left indicate the region being displayed. The left-most column shows the denoised version of the CTA slices. The natural CT noise (right column) appears to have similar local noise patterns to the generated Perlin noise (middle column).}
    \label{fig:noise}
\end{figure*}
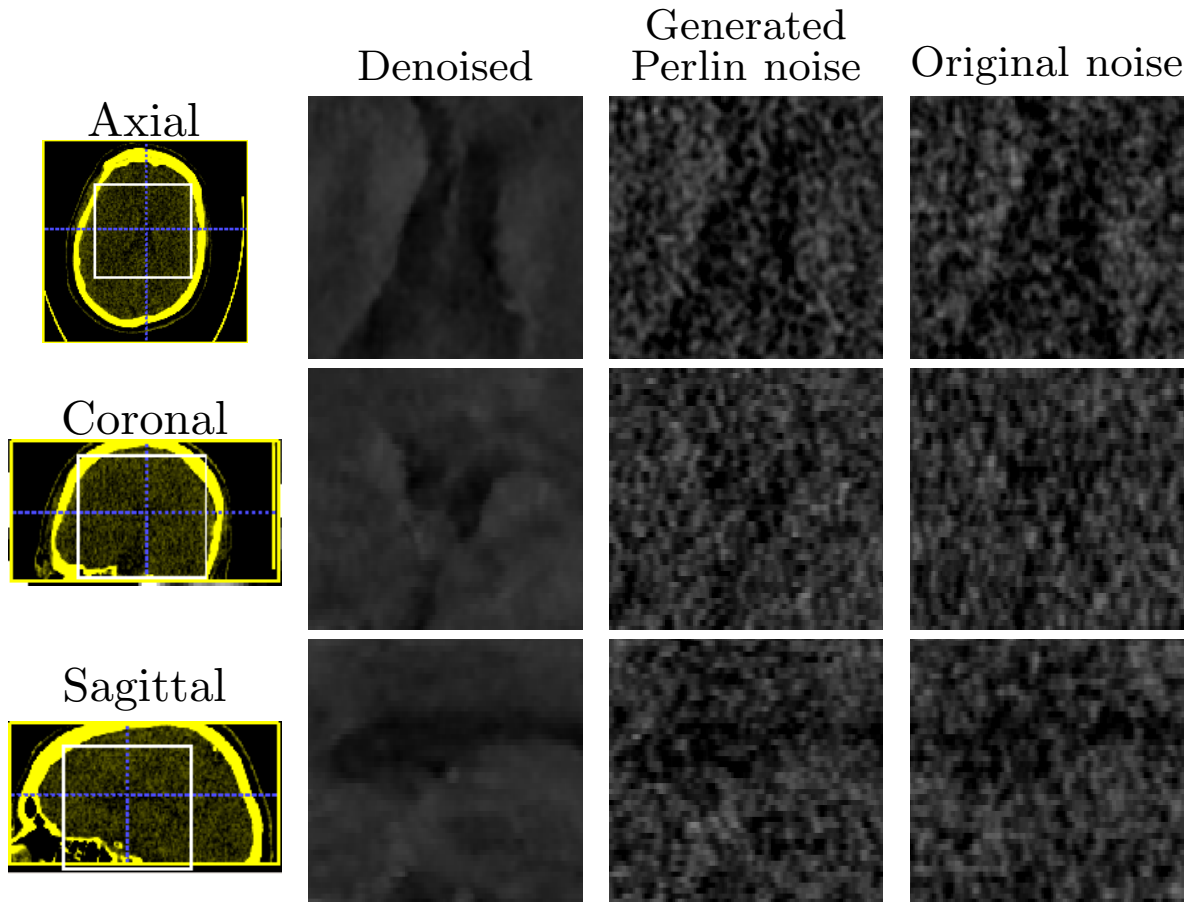
In order to avoid the pretraining phase to learn characteristics of individual patient's anatomy, we refrained from rendering the synthetic vessels into any of the scans in the labeled data set. To create the synthetic vessel data set the first time frame from each unlabeled patient was taken, since in those frames the contrast material had not yet reached the region of the scan.

Finally, both the synthetic vessel data set and the labeled data set were augmented through rotations, translations and mirroring. The amount of rotation and translation was performed by sampling a tuple of values from a uniform distribution. Rotations were in the range of $[-20, 20]$ degrees along the axial plane and $[-10, 10]$ along the coronal and sagittal planes. Translations were in the ranges $[-3\%, 3\%]$ of the scans vertical dimension along the z-axis and $[-5\%, 5\%]$ of the axial plane dimensions along the x and y axis. Mirroring only happened along the sagittal plane (with 50\% probability), as we aimed to keep the patient facing in the original direction and have the body retain the same orientation.

\subsection*{Synthetic Noise}

Recent studies have pointed out CNN approaches to image recognition are not as reliant on object shapes as was previously hypothesised, and instead these networks tend to develop large biases to textures present in the training data~\cite{cnn_textures}. Given that the CTA data set available to us is rather limited in number of examples (as is across most medical fields), it is not far-fetched to think that a CNN could learn patterns in the noise textures of the data set. We thus take the approach of denoising each 3D scan so as to later add new generated noise. The performance on the ´new noise´ data set is then compared to the model trained on scans where the original noise is left as is.

The denoising algorithm used was a variation of the `Non-local means` algorithm which, instead of using the mean of large pixel neighbourhoods of an individual image, the mean was computed based on the given pixel's value through time. The denoising process worked under the assumption that individual regions across the time dimension have independent noise from one-another. Denoised images such as the ones seen in Figure \ref{fig:noise} are obtained.

Extensive literature on noise in CTA scans exists~\cite{Omar_noise, Kijewski_1987, Polacin_1992}. Whenever a deeper understanding of underlying noise distributions is gained, better denoising and reconstruction algorihtm follow. Such studies however, attempt to identify approximate noise models by analysing inherent properties of noise distributions, most commonly the distances of individual pixel values to local regions of the image. Gaussian noise is often suggested as a fitting candidate. In computer graphics however, there is a clear distinction between ´value noise´ and ´gradient noise´ models. ´Value noise´ has the characteristic of lacking natural looking texture gradients. Despite this, the approximate distribution of pixel values on both of these types of noise models is practically identical.

Perlin noise~\cite{Perlin} is a form of ´gradient noise´ which shows resemblances to the observed textures in CTA scans while displaying global value distributions remarkably similar to Gaussian noise. In real 3D CTA scans pixel noise values are not independent of neighboring pixels. Noise gradients are consistent across neighboring 2D slices (regardless of orientation of the slices). The noise distribution can form local patterns that, at small enough scales, are practically indistinguishable from small brain vessels. To replicate this phenomenon, an approach similar to \cite{Perlin_data_augmentation} is used. A 3D volume of Perlin noise is added to the denoised scan in order to make the noise structure along each dimension consistent. We hypothesise that the consistency of the 3D noise across space might help the model better distinguish structures produced by noise.

\begin{figure}[!t]
  \begin{subfigure}{0.49\columnwidth}
    \includegraphics[width=\linewidth]{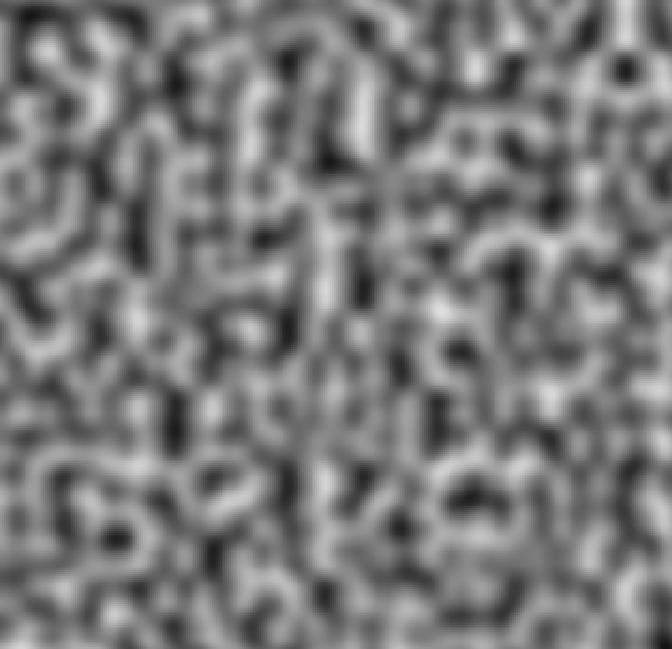}
  \end{subfigure}
  \hfill %%
  \begin{subfigure}{0.49\columnwidth}
    \includegraphics[width=\linewidth]{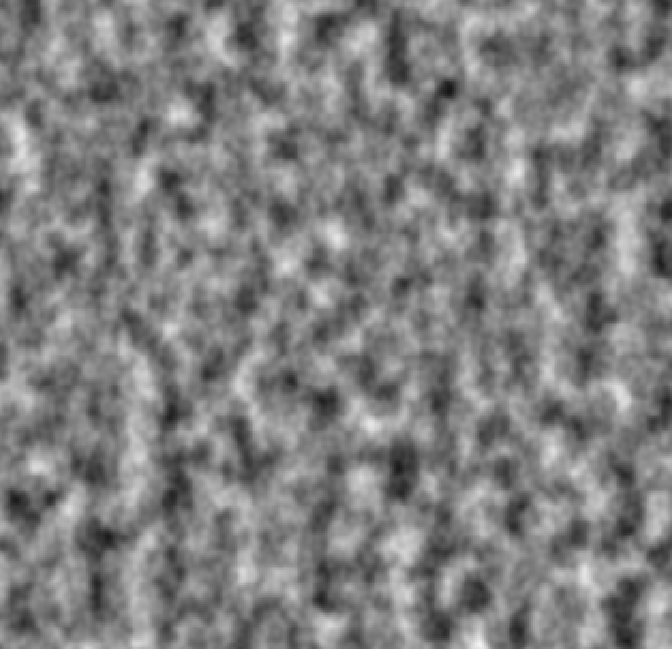}
  \end{subfigure}
  \caption{2D slices of the generated 3D Perlin noise. The single octave version is quite geometric in appearance (left), while the 3 octave version (right) has a more natural appearance.}
  \label{fig:octaves}
\end{figure}

It is also possible to generate this noise at multiple scales. Adding octaves of noise together can form fractal structures, which make the noise less geometric in appearance (see Figure~\ref{fig:octaves}). For this research we used 2 octaves. An example of the resulting scans can be seen in Figure~\ref{fig:noise}.

This addition of noise acts as a form of data-augmentation, from which we can generate any arbitrary number of new data points.

\subsection*{Network Architecture}
This choice of architecture for this research is a U-Net variant with similar parameters to the one posed by~\cite{DBLP:journals/corr/abs-1809-04430}. In their research, the architecture was tuned for a segmentation task of the neck and head anatomy. Despite the segmentation regions found in their tasks being significantly larger than the regions specific to the vessel segmentation task in ours, their architecture showed promise during initial tests.

The input to the network consists of 5 subsequent horizontal slices of a CTA scan. The network is trained to predict the ground truth for the middle slice out of the given 5. The 2 lower and 2 upper slices are thus meant to provide the network with useful context. The residual 3D U-net architecture consisted of 7 residual convolutional blocks in the downward path, having a fully-connected block at the bottom, and then followed by 7 residual convolutional block in the upwards path. The first 5 downwards layers performed 2D convolutions, while the latter 2 were 3D. As characteristic to U-Nets, the input to each upwards block was the output of the previous upwards block concatenated with the residual connection from the corresponding downwards block. The original x and y-dimensions of the input slices were preserved across all layers by using a stride of 1 and applying zero-padding where necessary. The output layer responsible for assigning a segmentation value to each pixel (in the original resolution of the input) was a 1x1x1 convolutional layer with sigmodial activation. Specific training hyperparameters used in this research can be found in the Appendix.

During training, a Dice-coefficient loss function was used. In preliminary experiments, this loss function proved to converge much faster and with much lower variance than a focal binary cross entropy loss function. To speed up convergence, the Adam optimizer \cite{adam} was used to compute adaptive learning rates for each parameter.

\section{Results, and Discussion}
\subsection*{Experimental Results}

When pre-training the model on generated vascular structures, 1000 unique scans were created. This was the case for both generation conditions (original CT noise and new Perlin noise). This translated to roughly 12500 training batches for each condition. None of the slices in the pre-training set were reused. In the other hand, when training and fine-tuning on real patient data, 5 epochs of iteration over the training set were performed. On a NVIDIA Titan RTX GPU, the pre-training procedure took roughly 16 hours per model, while the fine-tuning process was an additional 25 hours.

\begin{table*}[!t]
\caption{Average test set performances with corresponding standard deviations.}
\begin{tabular}{c|c|c|c|c|c|}
\cline{2-6}
          & \hfil Real data only & \begin{tabular}[c]{@{}l@{}}\hfil Pretrain\\\hfil original noise\end{tabular} & \begin{tabular}[c]{@{}l@{}}\hfil Fine-tune\\\hfil original noise\end{tabular} & \begin{tabular}[c]{@{}l@{}}\hfil Pretrain\\\hfil new noise\end{tabular} & \begin{tabular}[c]{@{}l@{}}\hfil Fine-tune\\\hfil new noise\end{tabular} \\ \hline
\multicolumn{1}{|l|}{\begin{tabular}[c]{@{}l@{}}\hfil Mean Dice\\ $\mu$ ($\pm$ $\sigma^ 2$)\end{tabular}} & \small 0.691 ($\pm$ 0.020) & \small 0.498 ($\pm$ 0.006) & \small 0.710 ($\pm$ 0.019) & \small 0.498 ($\pm$ 0.016) & \small 0.709 ($\pm$ 0.025)  \\ \hline
\end{tabular}
\label{Tab:results}
\end{table*}

\begin{figure}[!h]
    \centering
    \includegraphics[width=1.0\columnwidth]{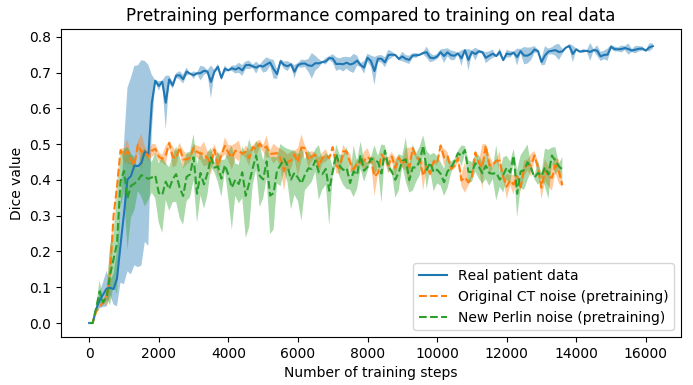}
    \caption{Dashed lines show validation Dice scores during pre-training process on generated vascular structures. For comparison, the solid line shows validation scores of the model trained exclusively on real images. Each line depicts the average across 3 separate training runs. The shaded area denotes one standard deviation from the mean.}
    \label{fig:pretrain_results}
\end{figure}

The results seen in Figure \ref{fig:pretrain_results} show the validation scores of the model trained on real patient scans side-by-side with the validation scores of models pre-trained on generated vascular structures. The results reported are averages across 3 training runs. Pre-training on generated scans where Perlin noise was applied appears to achieve similar maximum scores as models pre-trained on generated scans where the original CT noise was intact. This can also be seen in the test set scores in Table \ref{Tab:results}. The two pre-trained model conditions had no difference in Dice value in the test set, having a mean Dice value of 0.710 and 0.709. Meanwhile, the condition were models were not pre-trained had a mean Dice of 0.691.

\begin{figure}[!h]
    \centering
    \vspace{-0.5cm}
    \includegraphics[width=\columnwidth]{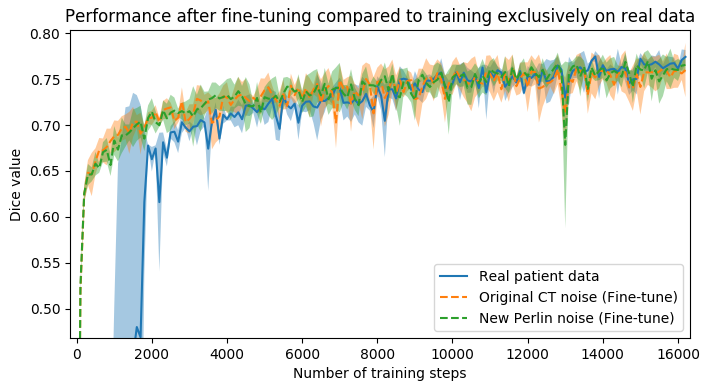}
    \caption{Dashed lines show validation Dice scores during fine-tuning processes. For comparison, the solid line shows validation scores of the model trained exclusively on real images. Each line depicts the average across 3 separate training runs.}
    \label{fig:finetune_results}
\end{figure}

Across all validation steps in each individual pre-training run, the best performing checkpoint in terms of Dice score was tracked. The model weights for this checkpoint were then used as a start point in the fine-tuning procedure. 

Figure \ref{fig:finetune_results} shows the validation scores of the fine-tuning progress alongside the ones obtained by the condition with no model pre-training. For the first 8000 training steps, the fine-tuning validation scores appear to surpass that of the model trained exclusively on real data by at least 0.02. This difference appears to diminish with subsequent training set iterations. Furthermore, the model pre-trained on synthetic vascular data with Perlin noise, appears to have a performance on par with the model pre-trained on scans that had the original CT noise untouched. The mean test set performances can be seen in Table \ref{Tab:results}.

\subsection*{Qualitative Results}
\begin{figure*}[!ht]
    \vspace{-0.5cm}
  \begin{subfigure}{.5\columnwidth}
    \includegraphics[width=\linewidth]{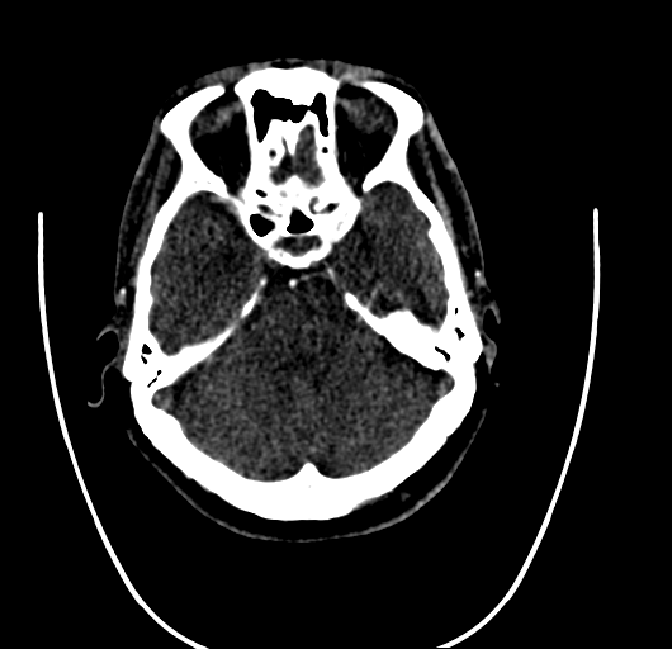}
  \end{subfigure}
  \hfill %%
  \begin{subfigure}{0.5\columnwidth}
    \includegraphics[width=\linewidth]{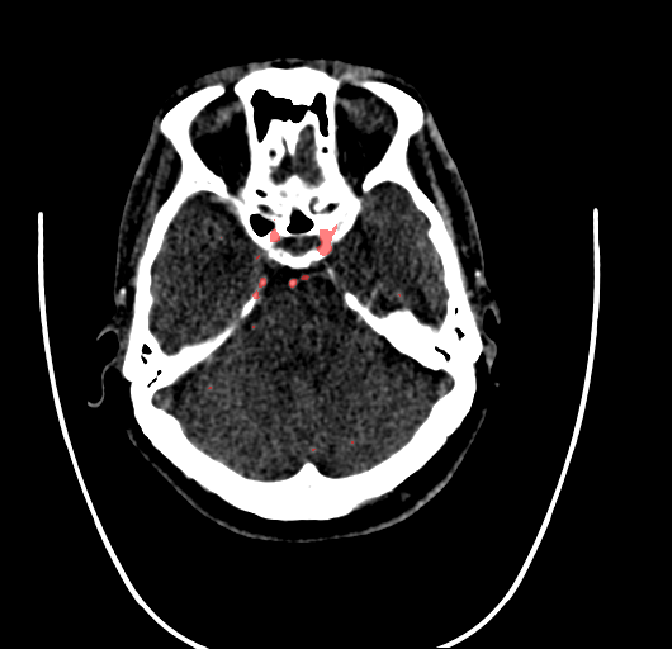}
  \end{subfigure}
   \hfill %%
  \begin{subfigure}{0.5\columnwidth}
    \includegraphics[width=\linewidth]{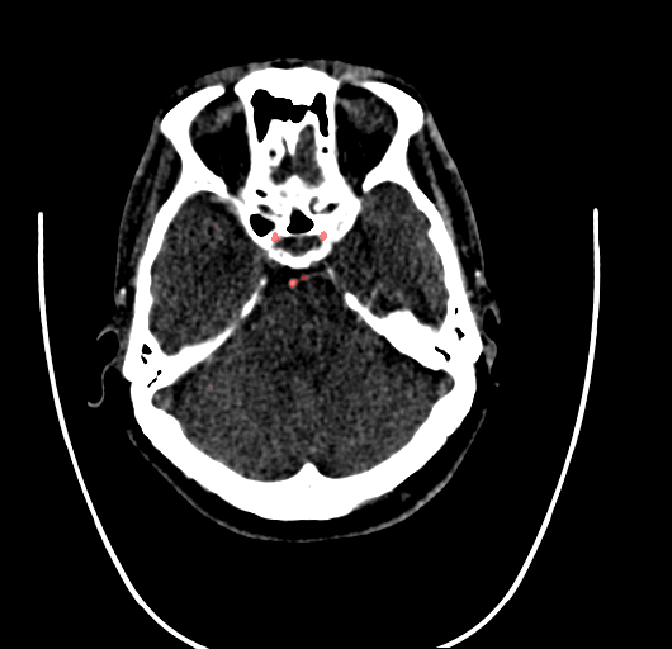}
  \end{subfigure}
    \hfill %%
  \begin{subfigure}{0.5\columnwidth}
    \includegraphics[width=\linewidth]{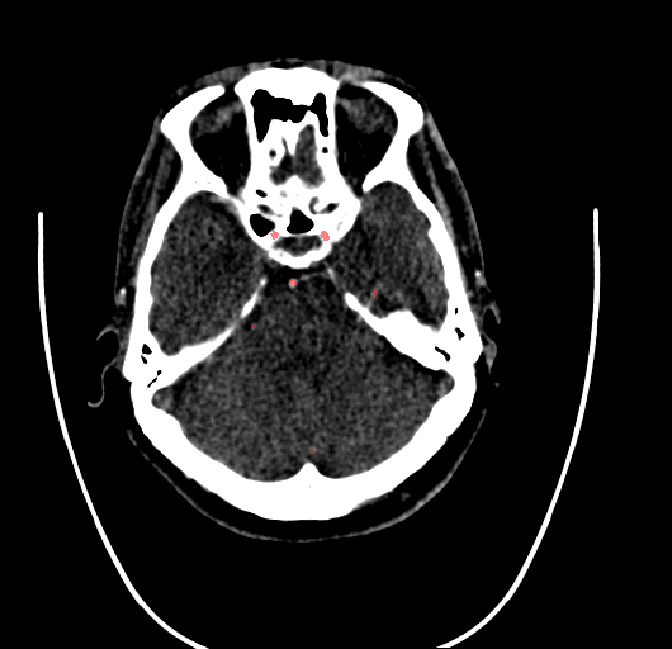}
  \end{subfigure}
  \caption{Segmentation on internal carotid arteries. Unaltered image (left). No pretrain model (mid-left). Fine-tuned model (mid-right). Hand-labeled ground truth (right).}
  \label{fig:segmentation1}
\end{figure*}

\begin{figure*}[!ht]
  \begin{subfigure}{.5\columnwidth}
    \includegraphics[width=\linewidth]{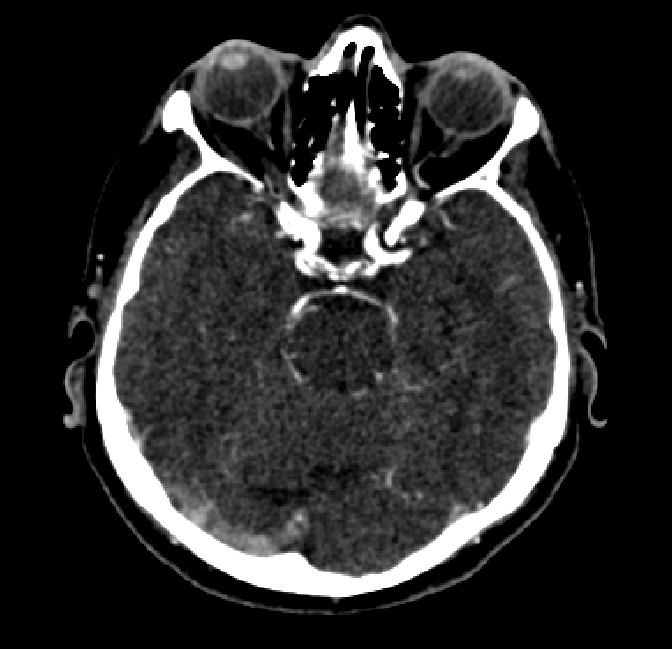}
  \end{subfigure}
  \hfill %%
  \begin{subfigure}{0.5\columnwidth}
    \includegraphics[width=\linewidth]{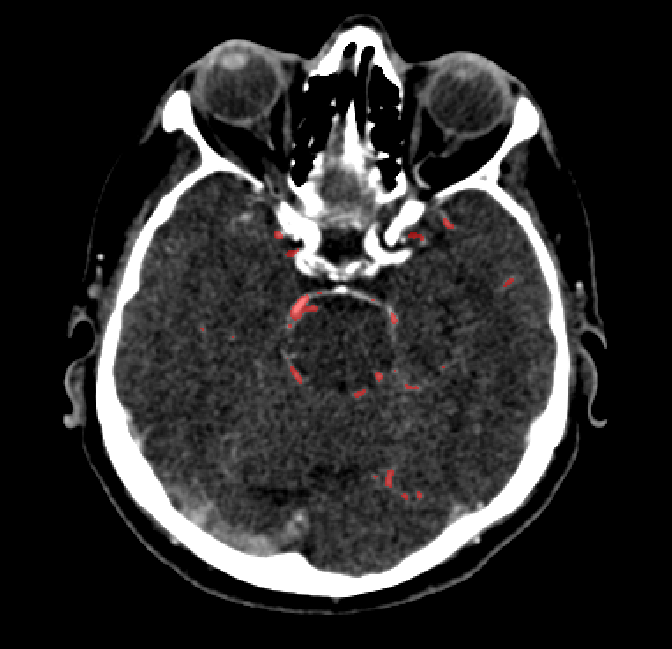}
  \end{subfigure}
   \hfill %%
  \begin{subfigure}{0.5\columnwidth}
    \includegraphics[width=\linewidth]{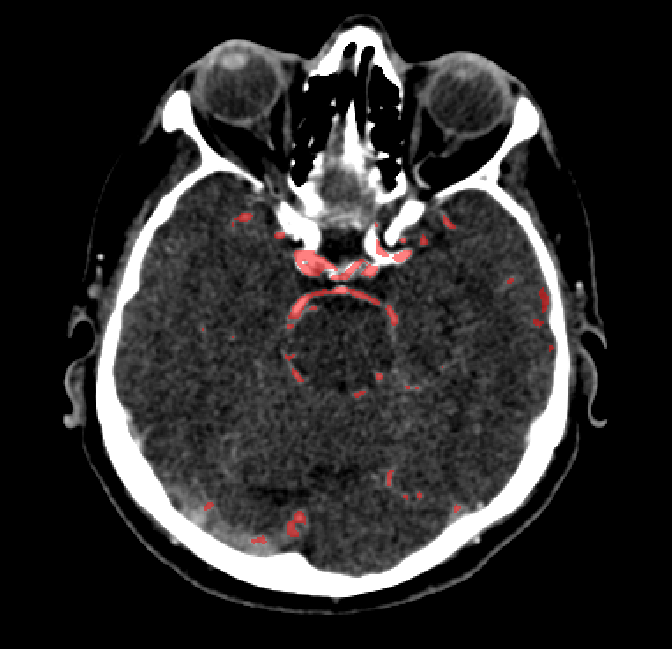}
  \end{subfigure}
    \hfill %%
  \begin{subfigure}{0.5\columnwidth}
    \includegraphics[width=\linewidth]{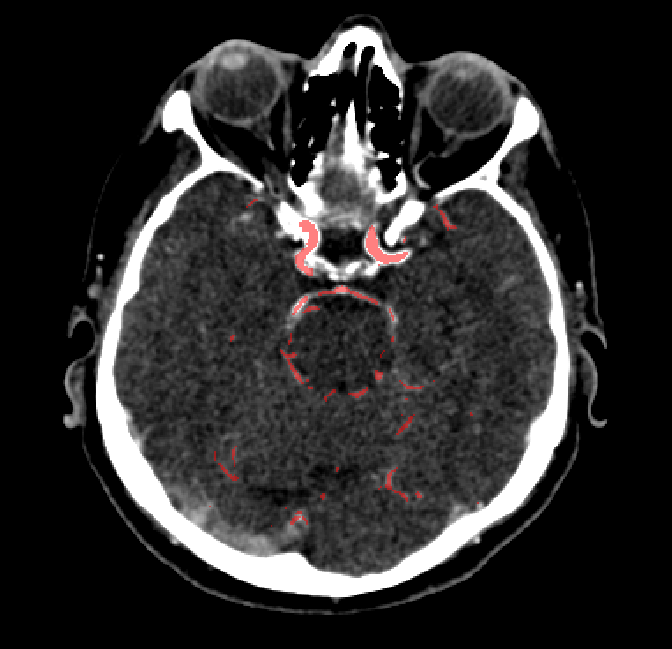}
  \end{subfigure}
  \caption{Segmentations on regions near internal carotid arteries. Unaltered image (left). No pretrain model (mid-left). Fine-tuned model (mid-right). Hand-labeled ground truth (right).}
  \label{fig:segmentation2}
\end{figure*}

\begin{figure*}[!ht]
  \begin{subfigure}{.5\columnwidth}
    \includegraphics[width=\linewidth]{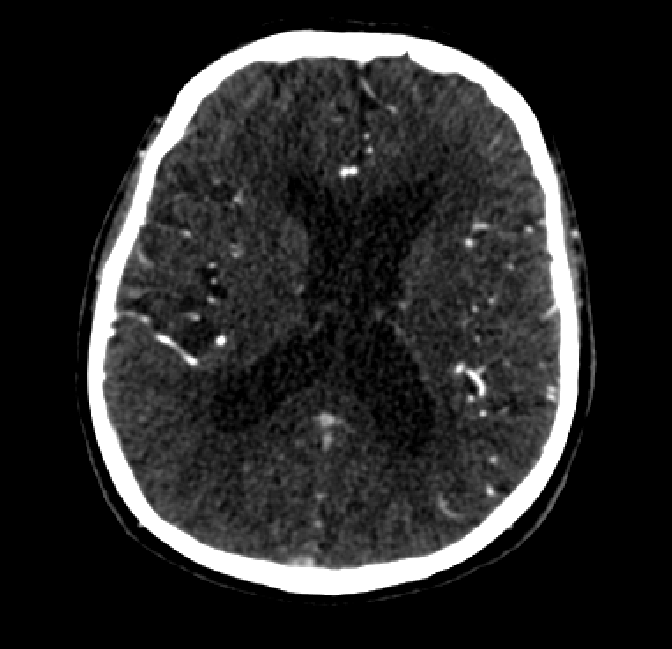}
  \end{subfigure}
  \hfill %%
  \begin{subfigure}{0.5\columnwidth}
    \includegraphics[width=\linewidth]{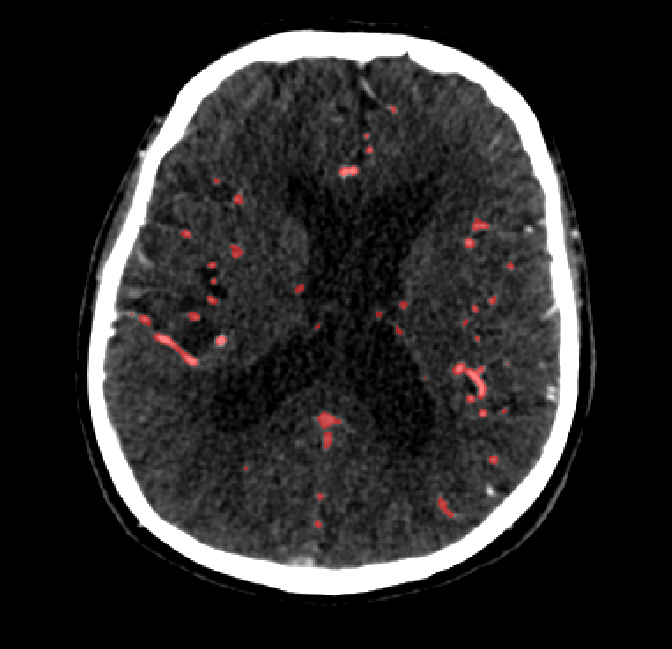}
  \end{subfigure}
   \hfill %%
  \begin{subfigure}{0.5\columnwidth}
    \includegraphics[width=\linewidth]{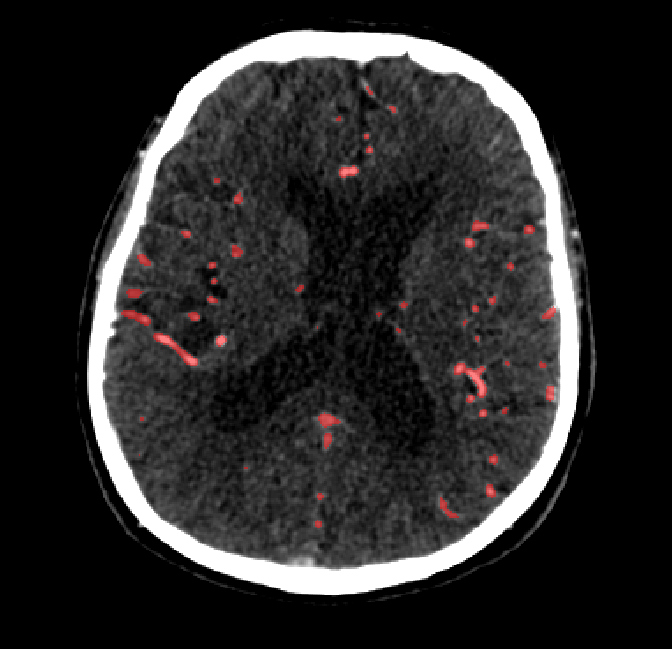}
  \end{subfigure}
    \hfill %%
  \begin{subfigure}{0.5\columnwidth}
    \includegraphics[width=\linewidth]{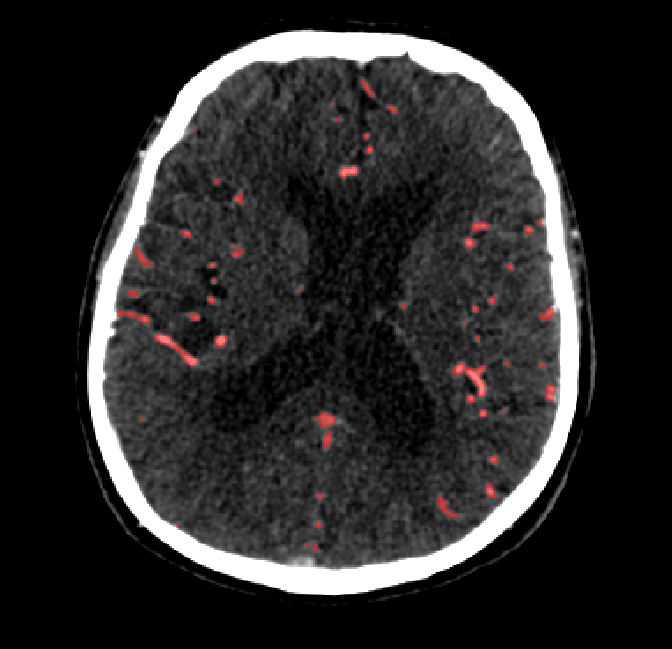}
  \end{subfigure}
  \caption{Segmentations on a central slice. Unaltered image (left). No pretrain model (mid-left). Fine-tuned model (mid-right). Hand-labeled ground truth (right).} 
  \label{fig:segmentation3}
\end{figure*}

\begin{figure*}[!ht]
  \begin{subfigure}{.5\columnwidth}
    \includegraphics[width=\linewidth]{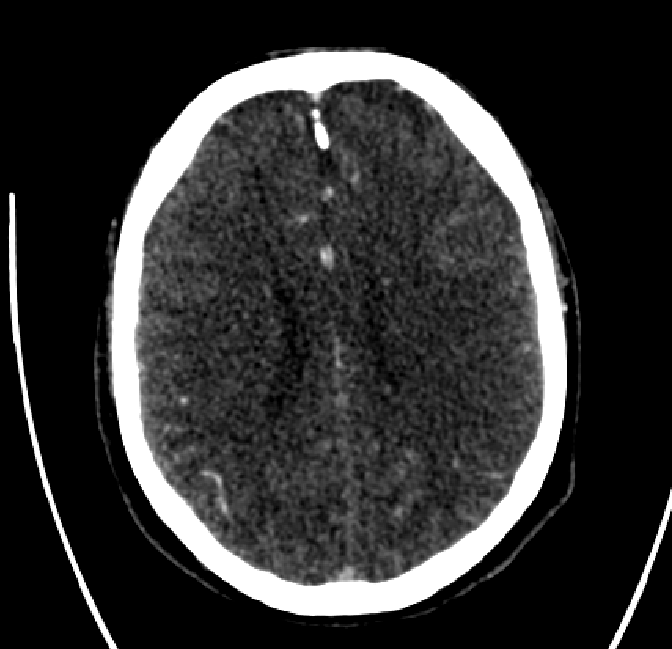}
  \end{subfigure}
  \hfill %%
  \begin{subfigure}{0.5\columnwidth}
    \includegraphics[width=\linewidth]{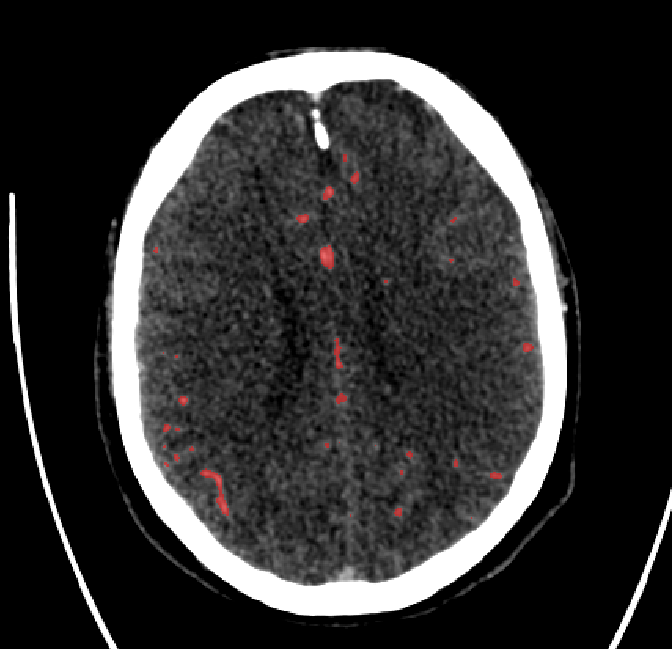}
  \end{subfigure}
   \hfill %%
  \begin{subfigure}{0.5\columnwidth}
    \includegraphics[width=\linewidth]{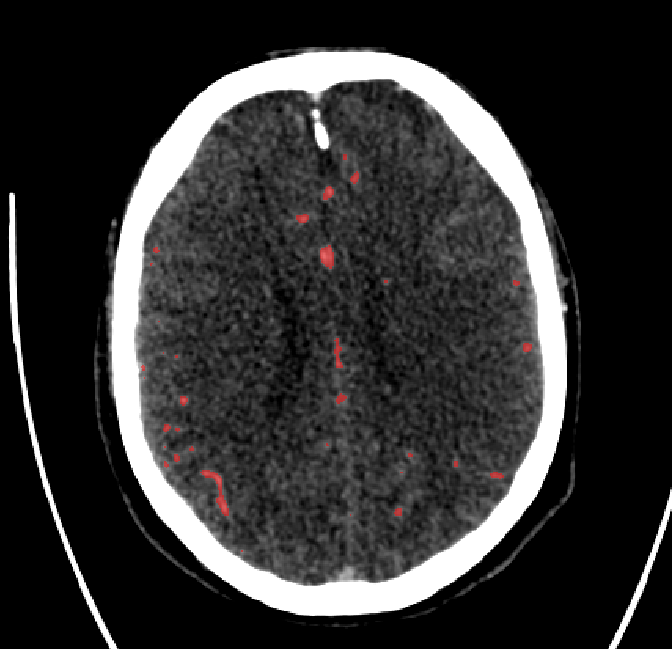}
  \end{subfigure}
    \hfill %%
  \begin{subfigure}{0.5\columnwidth}
    \includegraphics[width=\linewidth]{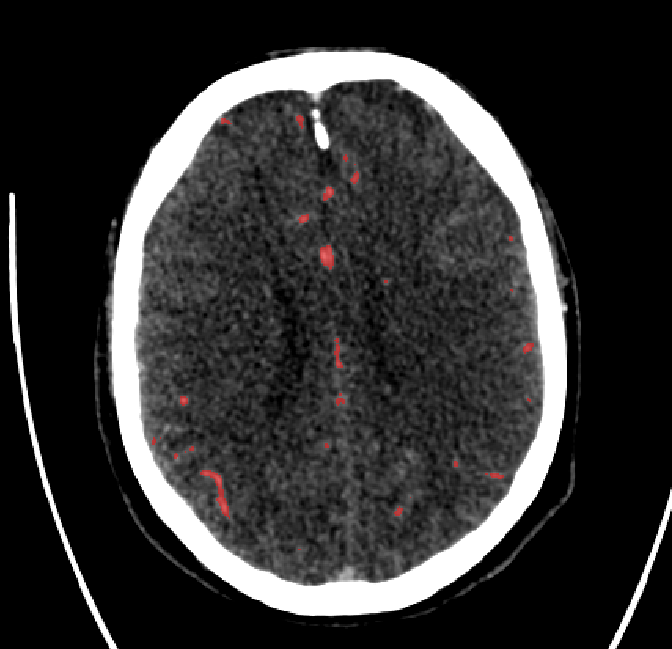}
  \end{subfigure}
  \caption{Segmentations on a central slice. Unaltered image (left). No pretrain model (mid-left). Fine-tuned model (mid-right). Hand-labeled ground truth (right).}
  \label{fig:segmentation4}
\end{figure*}

\begin{figure*}[!ht]
  \begin{subfigure}{.5\columnwidth}
    \includegraphics[width=\linewidth]{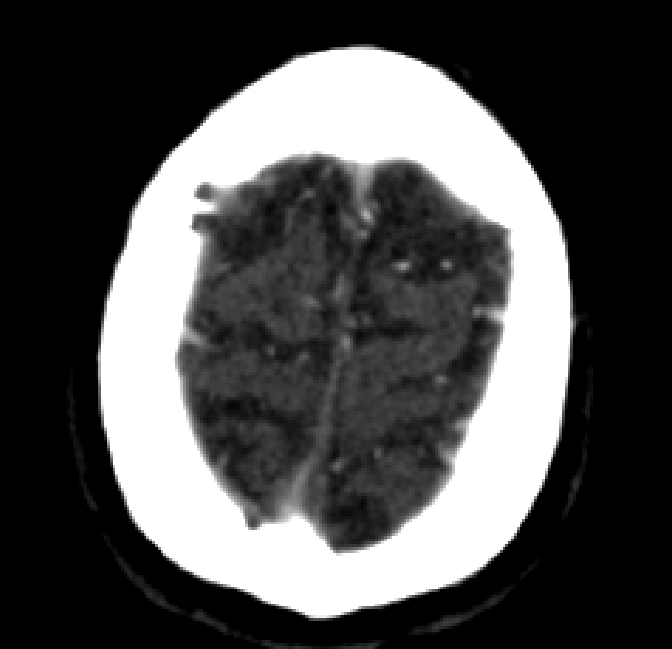}
  \end{subfigure}
  \hfill %%
  \begin{subfigure}{0.5\columnwidth}
    \includegraphics[width=\linewidth]{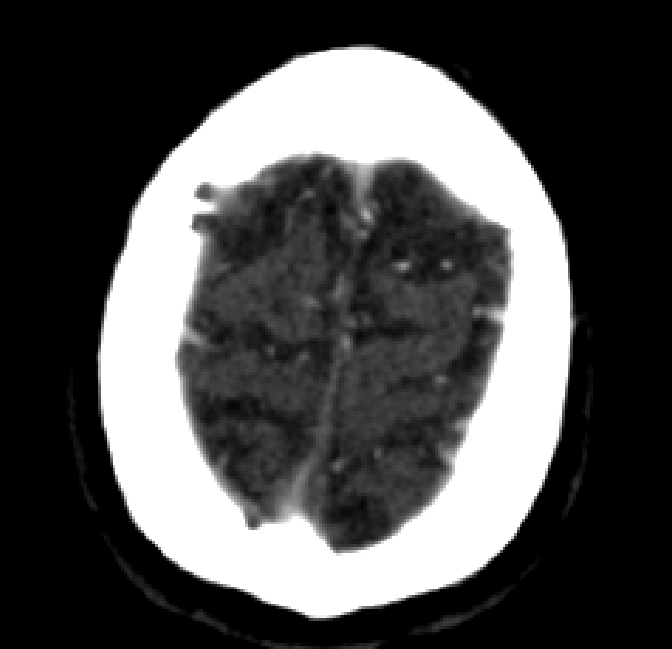}
  \end{subfigure}
   \hfill %%
  \begin{subfigure}{0.5\columnwidth}
    \includegraphics[width=\linewidth]{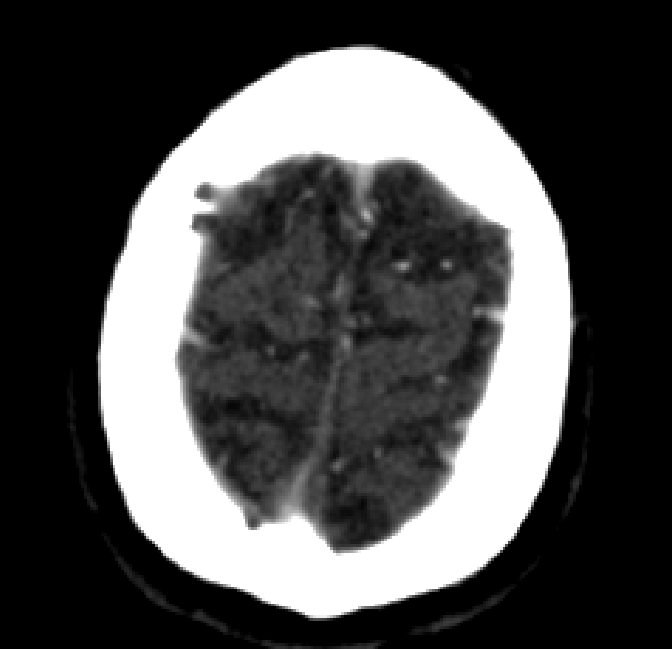}
  \end{subfigure}
    \hfill %%
  \begin{subfigure}{0.5\columnwidth}
    \includegraphics[width=\linewidth]{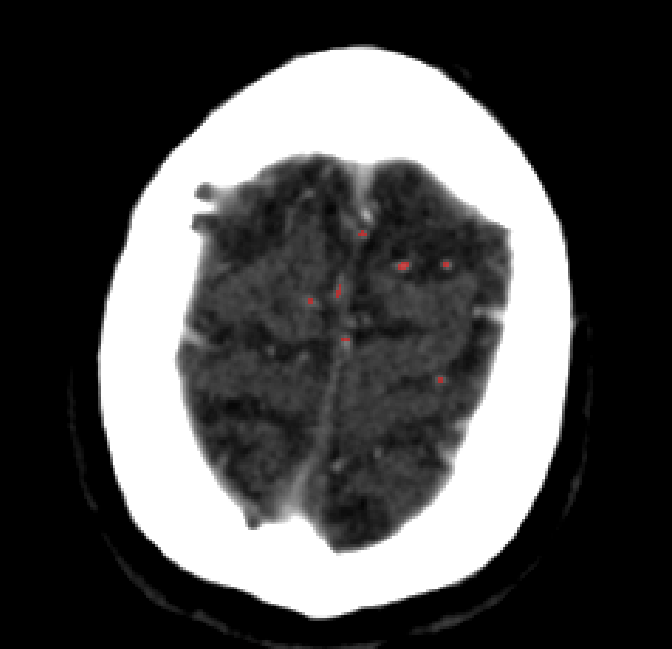}
  \end{subfigure}
  \caption{Segmentations on regions near the top of the skull. Unaltered image (left). No pretrain model (mid-left). Fine-tuned model (mid-right). Hand-labeled ground truth (right).}
  \label{fig:segmentation5}
\end{figure*}

\begin{figure*}[!ht]
  \begin{subfigure}{.5\columnwidth}
    \includegraphics[width=\linewidth]{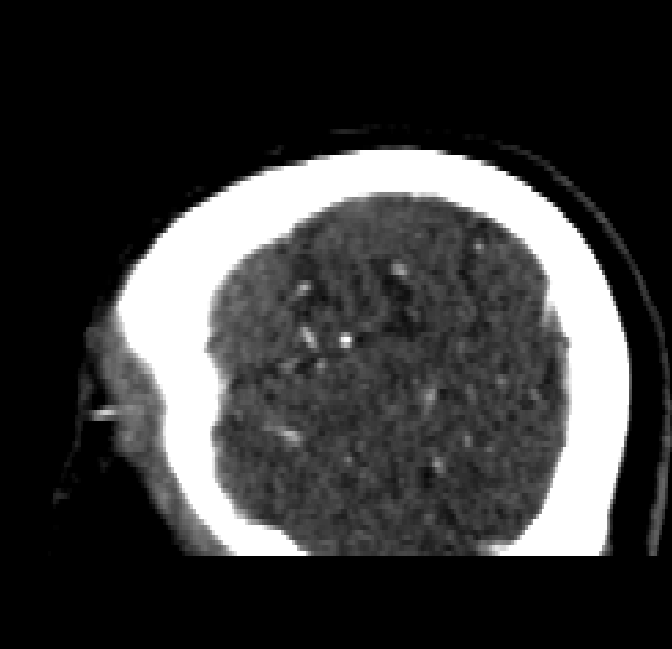}
  \end{subfigure}
  \hfill %%
  \begin{subfigure}{0.5\columnwidth}
    \includegraphics[width=\linewidth]{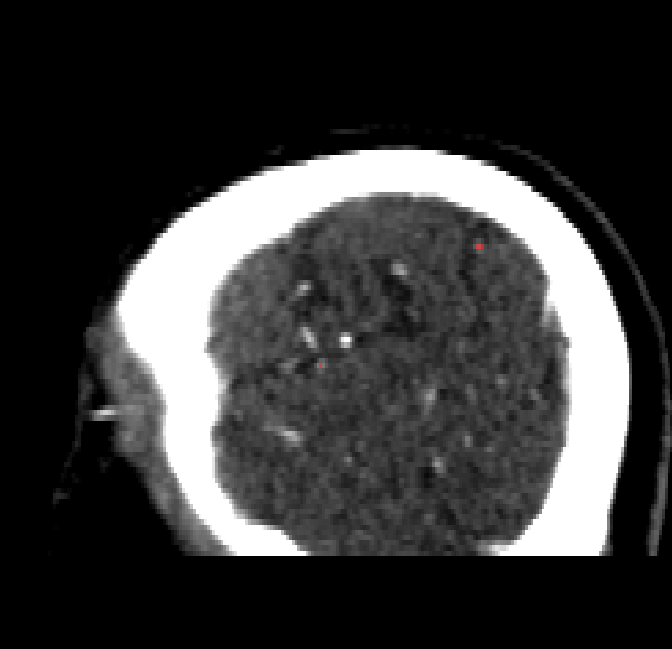}
  \end{subfigure}
   \hfill %%
  \begin{subfigure}{0.5\columnwidth}
    \includegraphics[width=\linewidth]{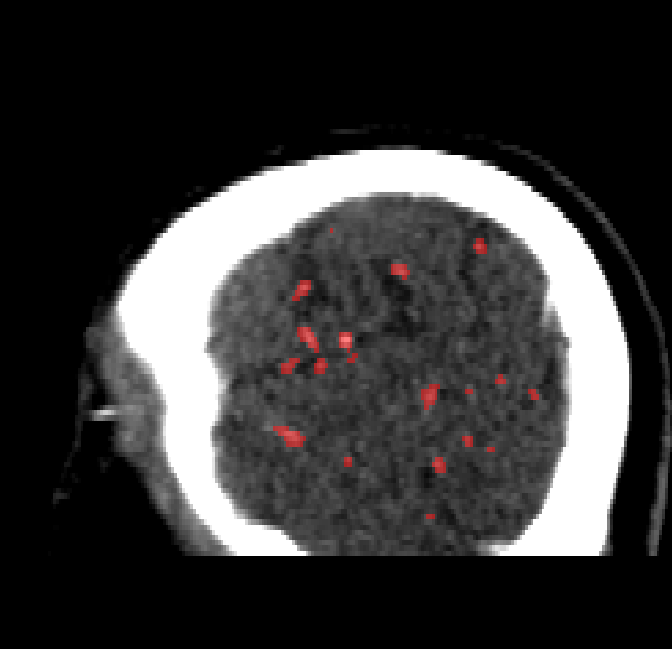}
  \end{subfigure}
    \hfill %%
  \begin{subfigure}{0.5\columnwidth}
    \includegraphics[width=\linewidth]{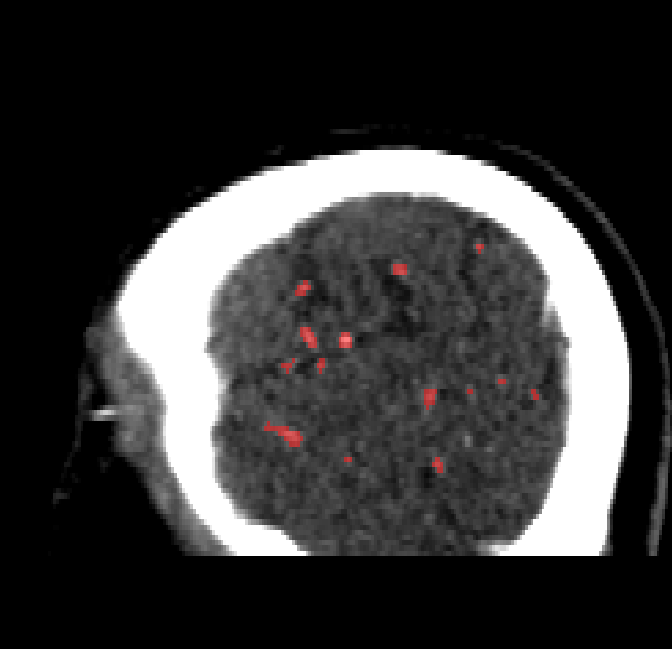}
  \end{subfigure}
  \caption{Side view of segmentations on regions near the left side of the skull. Unaltered image (left). No pretrain model (mid-left). Fine-tuned model (mid-right). Hand-labeled ground truth (right).}
  \label{fig:segmentation6}
\end{figure*}

\begin{figure*}[!ht]
  \begin{subfigure}{.5\columnwidth}
    \includegraphics[width=\linewidth]{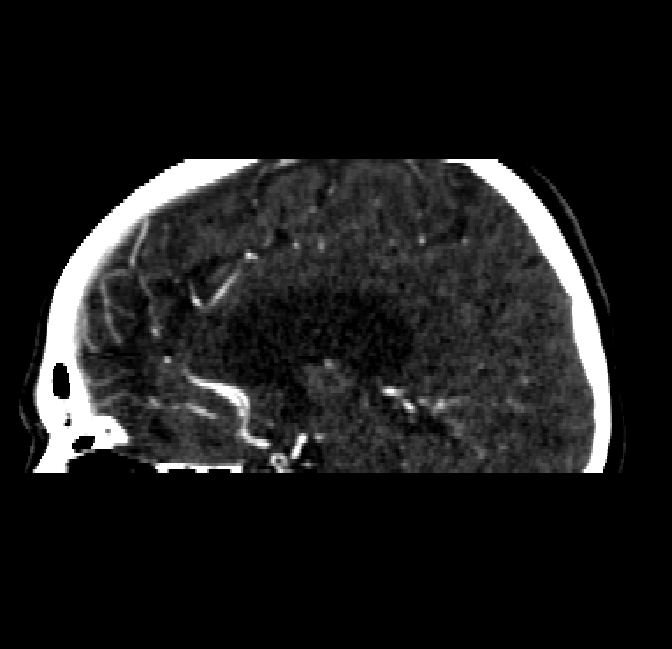}
  \end{subfigure}
  \hfill %%
  \begin{subfigure}{0.5\columnwidth}
    \includegraphics[width=\linewidth]{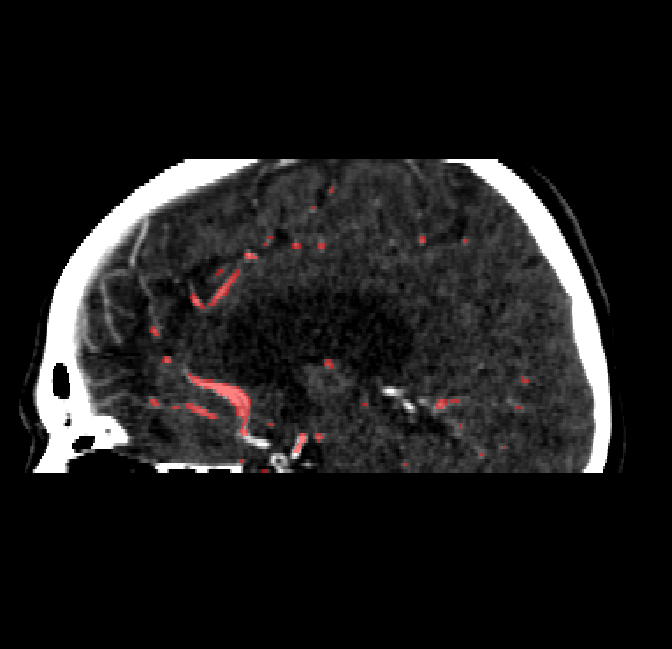}
  \end{subfigure}
   \hfill %%
  \begin{subfigure}{0.5\columnwidth}
    \includegraphics[width=\linewidth]{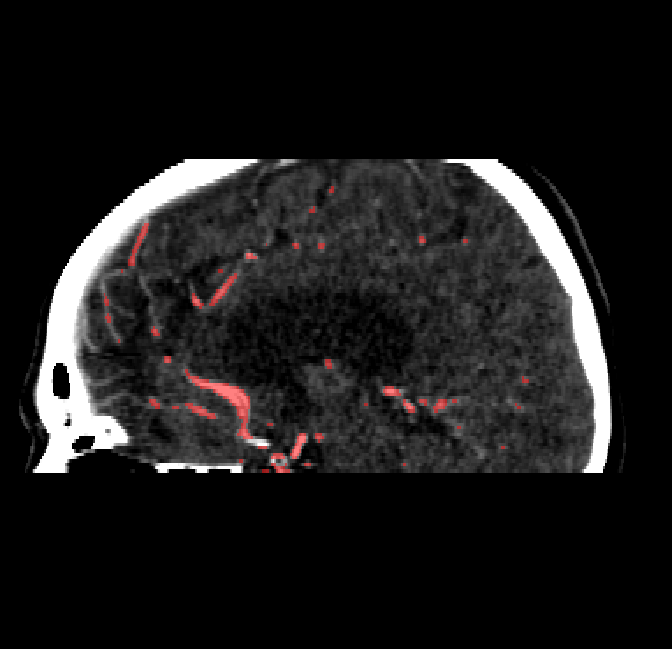}
  \end{subfigure}
    \hfill %%
  \begin{subfigure}{0.5\columnwidth}
    \includegraphics[width=\linewidth]{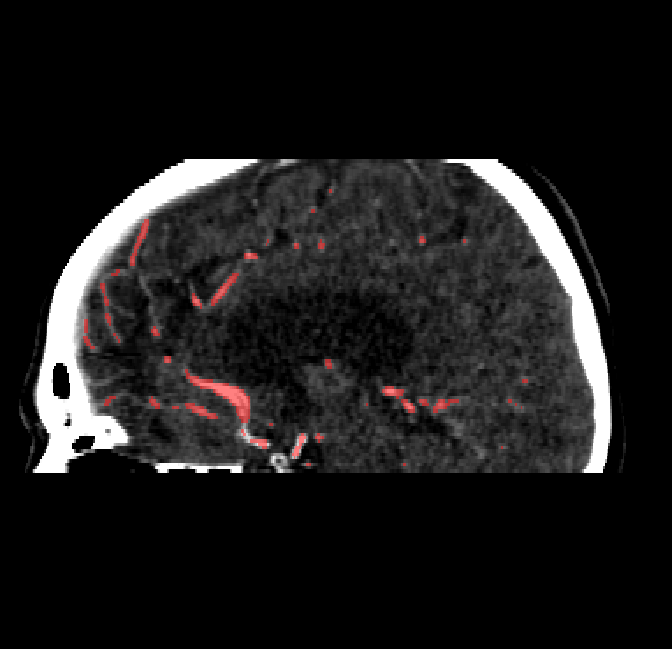}
  \end{subfigure}
  \caption{Side view of segmentations. Unaltered image (left). No pretrain model (mid-left). Fine-tuned model (mid-right). Hand-labeled ground truth (right).}
  \label{fig:segmentation7}
\end{figure*}

After fine-tuning, the two pre-train conditions appear to have little or no difference in terms of qualitative performance at the vessel segmentation task. However, when compared to the models trained only on real data, there are noticeable differences. Throughout Figures~\ref{fig:segmentation1}-\ref{fig:segmentation7}, the model pre-trained on scans with original CT noise is to be taken as representative of segmentation performed by the Perlin noise pre-training condition. Images showing the hand-labeled ground truth, as well as the unlabeled slice, are presented for comparison.

For vessels in the center of the head, all model types appear to accurately segment vessels. In certain cases, the models appear to learn to correctly avoid segmenting pieces of bone that could, in terms of shape and contrast, easily be confused with large vessels. An example of this can be seen in the frontal section of Figure~\ref{fig:segmentation4}.

The models appear to have a hard time segmenting vessels close to the skull surface. The model trained exclusively on patient data appears to struggle far more for these types on conditions than the fine-tuned models. Figure~\ref{fig:segmentation6} shows an example of the fine-tuneds model having close to no trouble segmenting vessels near the left side of the skull, while the baseline model suffers heavily from false negatives. To lesser degree, this effect can also be observed in the frontal lobe of Figure~\ref{fig:segmentation7} In the other hand, Figure~\ref{fig:segmentation5} presents an example of both models failing to segment vessels near the top of the skull. 

In the other hand, it should be noted that fine-tuned models suffered from false positives more often than models with no pre-training. The fine-tuned models appeared to occasionally segment regions near the skull, which although similar in intensity to vessels, had no resemblance in terms of shape. Examples of this are seen in Figure~\ref{fig:segmentation2} near the occipital bone and near the right temporal bone.
\clearpage
Examples of oversegmentation were observed to happen commonly around the internal carotid arteries. This is likely due to the amount of contact surface between the artery and the surrounding bone. There was also a tendency for all model conditions to segment bone regions that were similar in shape to large vessels (Figures~\ref{fig:segmentation1} and~\ref{fig:segmentation2}). Bone structures in such regions have similar pixel intensities to the arteries transporting contrast material, which could explain the source of confusion for a model. 

To our surprise, the models were able to occasionlly segment the shape of the internal carotid arteries correctly despite no boundary being visible to the naked eye between the vessel wall and the surrounding bone structure. An example of this can be seen in the fine-tuned model in Figure~\ref{fig:segmentation1}.

\subsection*{Discussion}

Even though the results show pretraining having relatively small improvements over the non-pretrained model, qualitative analysis showed notable differences between conditions in terms of false positives and false negatives. Is it thus hard to tell how much of the error is caused by each error type. Valuing one training condition over the other should be weighted by the intended use of the model.

As mentioned in the previous section, there are cases where the models fail to segment vessels close to the skull wall. However, we observed plenty of examples in the training set where such types of vessels were either not fully labeled, or not labeled altogether in the ground truth. Better labelling could help in reducing these sources of error.

Furthermore, segmentation of vessel structures such as the internal carotid arteries is a non-trivial task, even for a human. A lot of the vessel volume can be seen as easily segmentable through simple contrast features. However, correct segmentation of vessels around the base of the brain are likely to require complex hierarchical features that use surrounding structures. Data sets with larger varieties of patients could help supervised models develop a better understanding of such regions.

The models trained on scans where the CT noise was replaced with new Perlin noise provided no visible improvement over the other pretaining counterpart. One plausible argument is that the effect of overfitting to such reocurring noise structures in the repeated scans might simply be insignificant. Alternatively, it is possible that the rotations and translations performed during data augmentation (together with the interpolation process that comes with these) disrupt the reocurring noise patterns enough for them to not be an issue in the end.

Finally, when evaluating the generalizability of the presented results, it should also be noted that the CT scanner type and scanning protocol can vary across data sets. Although a variation in performance is expected under different technical factors of other CT scanners and image calibrations, the extent of the variation has not been explored in this study. 

\section{Conclusions}
High quality training data for vessel segmentation is scarce. There are two main reasons to this. The process of segmenting good ground truth training examples by hand is an expensive process in terms of the time required. Additionally, privacy laws make it hard for large patient data sets to become available for these kinds of research purposes.

In this paper we explored the use of synthetic vessel structures as a way to artificially increase the amount of training data. A computational model was created that generates volumetric vascular structures, which can then be blended into CT scans. 

Pre-training on such synthetic vascular structures and then fine-tuning on real data proved to improve the final Dice score by 0.019. The replacement of the original CT noise in the pre-training set had no impact on the performance. 

Qualitative analysis of the segmentation performed by fine-tuned models had notable differences from models trained exclusively on real data. While the regular (not pretrained) models had issues segmenting small vessels near the skull, the fine-tuned models proved capable of accurate segmentation of such regions. In the other hand, the fine-tuned models commonly displayed more false positives than the counterpart, specially on structures near the base of the brain.

\subsection*{Future Work}
As evident from this study, a downfall of computational vascular models is the difficulty to generate synthetic large-scale structures that conform, both visually and structurally, to those observed in CTA data. For purposes where realistic appearance is preferred over the explainability of the computational model, generative approaches similar to \cite{retina_GAN, retina_GAN_segmentation} could prove helpful under the condition that enough data is available. 

An investigation on the ability of generative approaches to create convincing detail could be of interest, specially at the level of the image noise distribution. Furthermore, the approach discussed in this paper of applying synthetic Perlin noise to an image does not have the capability to generate the artifacts in the noise pattern observed in real CT scans. Examples of such imperfections include beam hardening and scatter artifacts, as well as reconstruction artifacts. These could potentially be learned in the generative distributions.

The effect of large class imbalances common to volumetric imaging was also not investigated during this research. Addressing this issue could help improve segmentation in errors such as the oversegmentation of the skull around the carotid arteries or the false negatives on vessels near bone regions.

\subsection*{Acknowledgements}
Special thanks to Renan Sales Barros, Aashish Venkatesh, Efstratios Gavves, as well as the rest of the Nico.lab team, without whom this project would not have been possible.

\bibliographystyle{plain}
\bibliography{literature}

\clearpage
\section*{Appendix}
\begin{appendices}
    \scalebox{1}{
    \begin{tabular}{l|l}
        \textbf{General network parameters}                           & \textbf{Value}        \\ \hline
        Optimizer                                             & Adam                  \\
        Loss Function                                         & Dice-coefficient loss    \\
        Learning rate                                         & 0.00001              \\
        Weight Initializer                                    & Glorot Uniform        \\
        Hidden Layer Non-linearities                          & ReLU                 \\
        Batch normalization                                   & Yes                 \\
        Training Batch Size                                   & 8                    \\
        Validation Interval                                   & 100             \\
        Number Of Iterations Of Training Set                  & 5              \\
    \end{tabular}
    }
\end{appendices}

\end{document}